\input epsf
\input harvmac
\noblackbox
\newcount\figno
\figno=0
\def\fig#1#2#3{
\par\begingroup\parindent=0pt\leftskip=1cm\rightskip=1cm\parindent=0pt
\baselineskip=11pt
\global\advance\figno by 1
\midinsert
\epsfxsize=#3
\centerline{\epsfbox{#2}}
\vskip 12pt
\centerline{{\bf Figure \the\figno:} #1}\par
\endinsert\endgroup\par}
\def\figlabel#1{\xdef#1{\the\figno}}
\def\pano{\par\noindent}
\def\smno{\smallskip\noindent}
\def\meno{\medskip\noindent}

\font\cmss=cmss10
\font\cmsss=cmss10 at 7pt
\def\rlx{\relax\leavevmode}
\def\inbar{\vrule height1.5ex width.4pt depth0pt}
\def\IC{\relax\,\hbox{$\inbar\kern-.3em{\rm C}$}}
\def\IN{\relax{\rm I\kern-.18em N}}
\def\IP{\relax{\rm I\kern-.18em P}}
\def\ZZ{\rlx\leavevmode\ifmmode\mathchoice{\hbox{\cmss Z\kern-.4em Z}}
 {\hbox{\cmss Z\kern-.4em Z}}{\lower.9pt\hbox{\cmsss Z\kern-.36em Z}}
 {\lower1.2pt\hbox{\cmsss Z\kern-.36em Z}}\else{\cmss Z\kern-.4em Z}\fi}
\def\narrowplus{\kern -.04truein + \kern -.03truein}
\def\narrowminus{- \kern -.04truein}
\def\narrowminussub{\kern -.02truein - \kern -.01truein}

\def\si{\sigma}
\def\cl{\centerline}

\def\lb{\{ }
\def\rb{\} }

\def\type{type$\,{\rm II}\ $}
\def\typea{type$\,{\rm II}\,{\rm A}\ $}
\def\typeb{type$\,{\rm II}\,{\rm B}\ $}

\def\sqr#1#2{{\vcenter{\vbox{\hrule height.#2pt
            \hbox{\vrule width.#2pt height#1pt \kern#1pt
                  \vrule width.#2pt}\hrule height.#2pt}}}}


\lref\rbanks{ T.\ Banks, L.J.\ Dixon, D.\ Friedan and E.\ Martinec,
  {\it Phenomenology and conformal field theory, or Can string theory predict
  the weak mixing angle}?,
   Nucl.\ Phys.\ {\bf B299} (1988) 613}

\lref\rcls{ B.\ R.\ Greene and R.\ Plesser,
  {\it Duality in Calabi--Yau moduli space},
  Nucl.\ Phys.\ {\bf B338} (1990) 15\semi
  P.\ Candelas, M.\ Lynker and R.\ Schimmrigk,
  {\it Calabi--Yau manifolds in weighted $\IP_4$},
  Nucl.\ Phys.\ {\bf B341} (1990) 383}

\lref\rkv{S.\ Kachru and C.\ Vafa,
  {\it Exact results for $N=2$ compactifications of heterotic strings},
  Nucl.\ Phys.\ {\bf B450} (1995) 69, hep--th/9505105}
\lref\rkklmv{ S.\ Kachru, A.\ Klemm, W.\ Lerche, P.\ Mayr and  C.\ Vafa,
 {\it Nonperturbative results on the point particle limit of $N=2$ 
      heterotic string compactifications}, 
     Nucl.\ Phys.\ {\bf B459} (1996) 53, hep-th/9508155}

\lref\rcgh{P.\ Candelas, P.\ S.\ Green and T.\ H\"ubsch,
  {\it Finite distances between distinct Calabi--Yau vacua: (other
  worlds are just around the corner)},
  Phys.\ Rev.\ Lett.\ {\bf 62} (1989) 1956;
  {\it Rolling among Calabi--Yau vacua},
  Nucl.\ Phys.\ {\bf B330} (1990) 49}

\lref\rgms{A.\ Strominger,
  {\it Massless black holes and conifolds in string theory},
  Nucl.\ Phys.\ {\bf B451} (1995) 96, hep--th/9504090\semi
 B.\ R.\ Greene, D.\ R.\ Morrison and A.\ Strominger,
  {\it Black hole condensation and the unification of string vacua},
  Nucl.\ Phys.\ {\bf B451} (1995) 109, hep--th/9504145}

\lref\rbsw{R.\ Blumenhagen and A.\ Wi{\ss}kirchen,
  {\it Exactly solvable $(0,2)$ supersymmetric string vacua with GUT
  gauge groups},
  Nucl.\ Phys.\ {\bf B454} (1995) 561, hep--th/9506104\semi
 R.\ Blumenhagen, R.\ Schimmrigk and A.\ Wi{\ss}kirchen,
  {\it The $(0,2)$ exactly solvable structure of chiral rings,
  Landau--Ginzburg theories and Calabi--Yau manifolds},
  Nucl.\ Phys.\ {\bf B461} (1996) 460, hep--th/9510055 \semi
  R.\ Blumenhagen and A.\ Wi{\ss}kirchen,
  {\it Exploring the moduli space of $\,(0,2)$ strings},
  Nucl.\ Phys.\ {\bf B475} (1996) 225, hep-th/9604140}

\lref\rbsf{R.\ Blumenhagen, R.\ Schimmrigk and A.\ Wi{\ss}kirchen,
  {\it $(0,2)$ mirror symmetry},
  Nucl.\ Phys.\ {\bf B486} (1997) 598, hep--th/9609167 \semi
  R.\ Blumenhagen and S.\ Sethi,
  {\it On orbifolds of $(0,2)$ models},
  Nucl.\ Phys.\ {\bf B491} (1997) 263, hep--th/9611172 \semi 
  R.\ Blumenhagen and M.\ Flohr,
  {\it Aspects of  $(0,2)$ orbifolds and mirror symmetry},
  IASSNS-HEP-97/13,  hep--th/9702199}

\lref\rduala{ J.\  Distler and S.\ Kachru,
     {\it Duality of $(0,2)$ String Vacua},
     Nucl.\ Phys.\ {\bf B442} (1995) 64, hep-th/9501111}

\lref\rdualb{ T.-M.\ Chiang, J.\ Distler, B.\ R.\  Greene,
  {\it Some Features of $(0,2)$ Moduli Space},
       Nucl.\ Phys.\  {\bf B496} (1997) 590, hep-th/9702030}

\lref\rend{J.\  Distler, B.\ R.\ Greene, K.\ Kirklin and
           P.\ Miron,
    {\it Calculating endomorphism valued cohomology:
         Singlet spectrum in superstring models},
    Commun.\ Math.\ Phys.\ {\bf 122} (1989) 117\semi
    M.\ G.\ Eastwood and T.\ H\"ubsch,
    {\it Endomorphism valued cohomology and
         gauge neutral matter},
    Commun.\ Math.\ Phys.\ {\bf 132} (1990) 383\semi
    T.\ H\"ubsch, 
    {\it Calabi-Yau manifolds},
    World Scientific 1992\semi
    P.\ Griffiths and J.\ Harris,
   {\it Principles of algebraic geometry},
   John Willey \& Sons 1978}

\lref\rktb{ P.\ Berglund, C.\ V.\ Johnson, S.\ Kachru and 
            P.\ Zaugg,
       {\it Heterotic coset models and $(0,2)$ string vacua},
        Nucl.\ Phys.\ {\bf B460} (1996) 252, hep-th/9509170\semi
       M.\ Kreuzer and M.\ Nikbakht-Tehrani,
      {\it $(0,2)$ string compactifications},
      Talk at the International Symposium on the Theory
       of Elementary Particles, Buckow, August 27-31,
      hep-th/9611130 }

\lref\rew{E.\ Witten,
  {\it Phases of $N=2$ theories in two dimensions},
  Nucl.\ Phys.\ {\bf B403} (1993) 159, hep--th/9301042}

\lref\rkw{S.\ Kachru and E.\ Witten,
  {\it Computing the complete massless spectrum of a Landau--Ginzburg
  orbifold},
  Nucl.\ Phys.\ {\bf B407} (1993) 637, hep--th/9307038}

\lref\rdk{J.\ Distler and S.\ Kachru,
  {\it $(0,2)$ Landau--Ginzburg theory},
  Nucl.\ Phys.\ {\bf B413} (1994) 213, hep--th/9309110\semi
  J.\ Distler,
  {\it Notes on $(0,2)$ superconformal field theories},
   in Trieste HEP Cosmology 1994,hep--th/9502012}

\lref\rvitja{P.\ S.\ Aspinwall, B.\ R.\ Greene and D.\ R.\ Morrison,
    {\it Calabi-Yau moduli space, mirror manifolds and spacetime
     topology change in string theory},
    Nucl.\ Phys.\ {\bf B416} (1994) 414, hep--th/9309097\semi
  V.\ V.\ Batyrev,
  {\it Dual polyhedra and mirror symmetry for Calabi--Yau hypersurfaces
  in toric varieties},
  J.\ Alg.\ Geom.\ {\bf 3} (1994) 493, alg--geom/9310003}

\lref\rztwoinst{E.\ Witten,
{\it New issues in manifolds of $SU(3)$ holonomy},
   Nucl.\ Phys.\ {\bf B268} (1986) 79\semi
  M.\ Dine, N.\ Seiberg, X.\ Wen and E.\ Witten,
  {\it Nonperturbative effects on the string world sheet I,II},
  Nucl.\ Phys.\ {\bf B278} (1986) 769;
  {\it ibid.}\ {\bf B289} (1987) 319\semi
 J.\ Distler,
  {\it Resurrecting $(2,0)$ compactifications},
  Phys.\ Lett.\ {\bf B188} (1987) 431\semi
 J.\ Distler and B.\ R.\ Greene,
  {\it Aspects of $\,(2,0)$ string compactifications},
  Nucl.\ Phys.\ {\bf B304} (1988) 1}

\lref\rresolv{J.\ Distler, B.\ R.\ Greene and D.\ R.\ Morrison,
  {\it Resolving singularities in $(0,2)$ models},
   Nucl.\ Phys.\ {\bf B481} (1996) 289, hep--th/9605222\semi
   J.\ Distler, {\it private communication}}

\lref\rpri{J.\ Distler and S.\ Kachru,
  {\it Singlet couplings and $(0,2)$ models},
  Nucl.\ Phys.\ {\bf B430} (1994) 13, hep--th/9406090\semi
 E.\ Silverstein and E.\ Witten,
  {\it Criteria for conformal invariance of $\,(0,2)$ models},
   Nucl.\ Phys.\ {\bf B444} (1995) 161, hep--th/9503212}

\lref\rft{S.\ Sethi, C.\ Vafa, E.\ Witten,
      {\it Constraints on low-dimensional string compactifications},
      Nucl.\ Phys.\ {\bf B480} (1996) 213-224, hep-th/9606122 \semi
      I.\ Brunner, M.\ Lynker and R.\ Schimmrigk,\
      {\it F-theory on Calabi-Yau fourfolds},
     Phys.\ Lett.\ {\bf B387} (1996) 750, hep-th/9606148 \semi
    P.\ Mayr,
    {\it Mirror symmetry, $N=1$ superpotentials and tensionless 
     strings on Calabi-Yau  four-folds},
     Nucl.\ Phys.\ {\bf B494} (1997) 489, hep-th/9610162 \semi
    R.\ Friedman, J.\ Morgan and E.\ Witten,
    {\it Vector bundle and F-theory},
    IASSNS-HEP-97-3,  hep-th/9701162 \semi
    M.\ Bershadsky, A.\ Johansen, T.\ Pantev and V.\ Sadov,
    {\it Four dimensional compactification of F-theory},
    HUTP-96-A054, hep-th/9701165}

\lref\rcogp{ P.\ Candelas, X.\ C.\ De La Ossa, P.\ S.\ Green 
             and  L.\  Parkes,
        {\it A pair of calabi-Yau manifolds as an exactly soluble
             superconformal theory},
      Nucl.\ Phys.\ {\bf B359} (1991) 21}

\lref\rschu{ S.\ Katz and S.\ A.\ Str\o mme,
         {\it Schubert: a maple package for intersection
       theory}\semi
        J.\ DeLoera,
      {\it Puntos} }
     
\Title{\vbox{\hbox{hep--th/9707198}
                 \hbox{IASSNS--HEP--97/86}}}
{Target Space Duality for (0,2) Compactifications}
\smallskip
\centerline{{Ralph Blumenhagen${}^1$}   }
\bigskip
\centerline{${}^{1}$ \it School of Natural Sciences,
                       Institute for Advanced Study,}
\centerline{\it Olden Lane, Princeton NJ 08540, USA}
\smallskip
\bigskip
\bigskip\bigskip
\centerline{\bf Abstract}
\noindent
The moduli spaces of two $(0,2)$ compactifications of the heterotic
string can share  the 
same Landau-Ginzburg model even though at large radius they 
look completely different. It was argued that such a pair of 
$(0,2)$ models might be connected via a perturbative transition 
at the Landau-Ginzburg point.  
Situations of this kind are studied for 
some explicit models. By calculating the exact dimensions of
the generic moduli spaces at large radius, 
strong indications are found in favor
of a different scenario. The two moduli spaces are isomorphic 
and complex, K\"ahler and bundle moduli get exchanged. 
\footnote{}
{\pano
${}^1$ e--mail:\ blumenha@sns.ias.edu
\pano}
\Date{07/97}
\newsec{Introduction}

String compactifications to four dimensions with $N=1$ supersymmetry are
the most promising class of models to describe our physical universe. 
For the heterotic string the condition of $N=1$ supersymmetry leads one 
to models with $(0,2)$ supersymmetry from the world sheet perspective \rbanks.
This class of models has been the subject of study all over the last 
decade \refs{\rztwoinst\rdk\rduala\rpri\rbsw\rresolv\rdualb\rktb{--}\rbsf}.
However, especially during recent years not only the 
left-right symmetric subclass of $(2,2)$ supersymmetric models  was under
investigation but also new results were obtained establishing 
more general $(0,2)$ supersymmetric models. Former doubts about the 
consistency of such models could be shown to be not justified at least
for the class of $(0,2)$ models described by linear $\si$-models \rpri. 
Furthermore, exactly solvable $(0,2)$ models were constructed and identified
with some special points in the moduli space of such linear  $\si$-models,
showing directly that these models are consistent \rbsw. 

For the class of $(2,2)$ models the target space duality commonly known
as mirror symmetry turned out to be of primary importance to solve the model
on the K\"ahler moduli space \refs{\rcls\rcogp{--}\rvitja}. 
These results were used later to establish
a new non-perturbative duality between models with $N=2$ supersymmetry, namely
between the \typea  string on a $K3$ fibered Calabi-Yau threefold and
the heterotic string on $K3\times T_2$ \rkv. 
The local analysis of this stringy 
duality reproduced the results of Seiberg/Witten on the Coloumb moduli space
of $N=2$ gauge theories \rkklmv. 
In \rbsf\ the notion of mirror symmetry was generalized
to $(0,2)$ models, by showing that some of the mirror constructions
known from the $(2,2)$ context, like orbifolding, can be carried over to the 
$(0,2)$ case. 

F-theory on elliptically fibered Calabi-Yau fourfolds provides 
another huge class of consistent $N=1$ supersymmetric models in four 
dimensions \rft. These models are defined as \typeb  compactifications on
so-called D-manifolds, which are not necessarily Ricci-flat and support
D-branes on some submanifolds. If the fourfold
is also $K3$ fibered then such a compactification is expected to be dual 
to an elliptically fibered  heterotic $(0,2)$ model with in general 
a number of five-branes wrapped around the toroidal fiber. 

The subject of this paper is another duality holding in the class of 
$N=1$ models. It was first observed in \rduala\  and further elaborated in 
\rdualb\ that two at first sight different $(0,2)$ models can share 
the same Landau-Ginzburg 
model at small radius. Since in general the number of gauge singlets at
the Landau-Ginzburg point was known to be bigger than in the large radius
geometric phase one could imagine a situation similar to a 
conifold transition \rgms.
One starts at large radius on one $(0,2)$ model, moves down to small radius,
hits the Landau-Ginzburg point, turns on one of the new moduli and finally
finds oneself on the other $(0,2)$ model. 
The aim of this paper is to argue
that such a transition is very unlikely to happen and can be excluded
definitely in one of our examples involving the well known quintic. 
Technically, what will be done is to calculate the dimension of the three
relevant cohomology classes $H^1(M,T)$, $H^2(M,T)$ and $H^1(M,{\rm End}(V))$
giving complex, K\"ahler and bundle moduli of the $(0,2)$ compactification,
respectively. The result is that in all studied examples 
the dimension of the large radius geometric moduli space for two dual models 
agrees. This is so, even if the Landau-Ginzburg model provides  additional
singlets. 
This result points not to a transition to occur at the Landau-Ginzburg point,
but instead to an actual target space duality between different $(0,2)$
moduli spaces, where complex, K\"ahler and bundle moduli get exchanged.
The possibility of an isomorphism of moduli spaces was mentioned before in 
\rdualb, where a third scenario was also taken into account. 
Two models could be related
by something similar to a flop transition, thus describing an overall
model in different regions of its moduli space. Since the whole 
bundle moduli space of the quintic is visible in the parameters of the
linear $\si$-model, it can be compactified. Consequently,
one does not have any boundaries, so that the possibility of a flop like 
transition will be excluded in the following discussion. 

This paper is organized as follows. In section two some facts about linear 
$\sigma$-models and the duality at the Landau-Ginzburg point are briefly 
reviewed. In section three  one example is discussed in very much detail. 
It deals with the quintic and a dual $(0,2)$ candidate, for which the 
bundle valued cohomology classes are calculated. 
The technical aspects are discussed fairly  explicitly, not to bother
the reader with some boring technical aspects but to provide some further
compressed  reference for such calculations. The available techniques 
are scattered around the literature and are most often limited to ordinary
projective spaces \rend. Whereas, here one has to deal with more general
toric varieties.
In section four the same techniques are applied to some further examples
involving the sextic $\IP_{1,1,1,1,2}[6]$, for which two dual 
$(0,2)$ models are found featuring the same dimension of the geometric 
moduli spaces.

\newsec{Linear $\sigma$-models}

The primary reason $(0,2)$ models have become accessible to study in recent 
times is the development of the gauged linear $\si$-model by Witten \rew. This 
model is a relatively tractable massive two-dimensional field theory which is 
believed, under suitable conditions, to flow in the infrared to a non-trivial 
superconformal field theory. One of the more interesting features of the 
linear 
$\si$-model is its various connected vacua, or phases. At low energies, these 
phases appear to correspond to theories such as a non-linear  $\si$-model, a 
Landau-Ginzburg orbifold, or some other more peculiar theory like a hybrid 
model. The linear $\si$-model provides a natural setting in which the relation 
between some of these various types of theories can be studied.  

Let us begin by describing the fields in the $(0,2)$ linear  $\si$-model.
To shorten the notation it is assumed in the following review
that there is only one $U(1)$ gauge field. The generalization to
more gauge fields is straightforward and can be found in 
\refs{\rresolv,\rew}.  There 
are two sets of chiral superfields:  $\lb X_i\vert i=1,\ldots,N_x\rb$ with 
$U(1)$ charges
$\omega_i$ and $\lb P_l\vert l=1,\ldots,N_p\rb$ with $U(1)$ charges $-m_l$.
Furthermore, there are two sets of Fermi superfields: $\lb\Lambda^a\vert 
a=1,\ldots,N_{\Lambda}\rb$ 
with charges $n_a$ and $\lb\Gamma^j\vert j=1,\ldots,N_{\Gamma}\rb$ with 
charges $-d_j$.
The superpotential of the linear $\si$-model is given by,
\eqn\superpot{ S=\int d^2 z d\theta \left[ \Gamma^j W_j(X_i) + 
               P_l \Lambda^a F_a^l(X_i)  \right], }
where $G_j$ and $F_a^l$ are quasihomogeneous polynomials whose degree is 
fixed by 
requiring charge neutrality of the action. To ensure the absence of gauge 
anomalies  the following conditions have to be satisfied:
\eqn\anfree{\eqalign{  &\sum \omega_i = \sum d_j, \cr & \sum n_a = 
                       \sum m_l, \cr
                       &\sum d_j^2 - \sum w_i^2 = \sum m_l^2 - \sum n_a^2. \cr 
}}
If there is more than  one $U(1)$ gauge field the linear  conditions have to be
satisfied for every single $U(1)$  and the quadratic condition for every
pair of $U(1)$s. Thus, $N$ different $U(1)$ symmetries  give rise to $N(N+1)/2$
quadratic conditions.

In the large radius limit $r\gg 0$,  the model 
describes a $(0,2)$  non-linear $\si$-model on a generally singular
weighted projective space,
$\IP_{\omega_1,\ldots,\omega_{N_x}}[d_1,\ldots,d_{N_{\Gamma}}]$, 
with a coherent sheaf 
of rank $N_{\Lambda}-N_p-N_F$ defined as the cohomology of the monad
\eqn\sheaf{  0\to\ \bigoplus_{i=1}^{N_F} {\cal O} 
                {\buildrel \otimes E_a^i \over  \to}
               \bigoplus_{a=1}^{N_{\Lambda}}{\cal O}(n_a)
              {\buildrel \otimes F_a \over \to}
              \bigoplus_{l=1}^{N_p}{\cal O}(m_l)\to0.}
Here $N_F$ additional fermionic gauge symmetries have been introduced 
to make the model consistent. For an extended discussion of these 
fermionic gauge symmetries take a look into \rdualb. 
In the subsequent sections the notation 
\eqn\notat{ V(n_1,\dots,n_{N_{\Lambda}};m_1,\dots,m_{N_p}) \to
         \IP_{\omega_1,\dots,\omega_{N_x}}[d_1,\ldots,d_{N_{\Gamma}}] }
will be used for the singular configuration. 
For the situation where $N_{p}=1$ and $r\ll 0$, the low-energy physics is 
described by a Landau-Ginzburg orbifold with a superpotential
\eqn\lgsupo{    W(X_i,\Lambda^a,\Gamma^j)= \sum_j \Gamma^j W_j(X_i) 
                      + \sum_a \Lambda^a F_a(X_i). }
It was first observed in \rduala\  that in this superpotential the constraints 
$G_j$ and $F_a$ appear on equal footing, so that in particular
an exchange of them does not change the Landau-Ginzburg model as long as
all anomaly cancellation conditions are satisfied. In \rdualb\ this duality was
further elaborated showing that this exchange is still possible after
resolving the generically singular base manifold. The fact that at the
Landau-Ginzburg point the number of gauge singlets is usually bigger than 
in the geometric large radius limit was viewed as a hint  that 
a transition from  one $(0,2)$ model to another one takes place right
at the Landau-Ginzburg point. This transition would be a perturbative 
analogue of the non-perturbative conifold transition for \type  models \rgms. 
It will be shown in the next section that at least for some simple examples
such a scenario does not survive more refined tests.

\newsec{The quintic and its (0,2) dual}

The quintic is probably the most studied example of a Calabi-Yau 
compactification in the literature. 
It is given by a quintic hypersurface $M$ in the projective
space $\IP_4$. The bundle $V_M$ over $\IP_4[5]$ is a deformation of
the tangent bundle and can be  described as the cohomology of the monad
\eqn\qbundle{ 0\to\ {\cal O}\vert_{M}\to\bigoplus_{a=1}^{5}{\cal O}(1)\vert_{M}
\to{\cal O}(5)\vert_{M}\to0.}
The notation ${\cal O}(n)\vert_{M}$ means the line bundle on the ambient 
space with first Chern class $n\eta$ restricted to the hypersurface $M$. 
What is needed  for the following is the dimension of the moduli space.
At large radius one finds $h^1(M,V_M)=101$ complex deformations, 
$h^2(M,V_M)=1$ K\"ahler deformations and $h^1(M,{\rm End}(V_M))=224$ 
bundle deformations adding up to a 
total dimension of 326. In this particular case 
one finds the same number of moduli at the Landau-Ginzburg point \rkw. 
Moreover, E. Silverstein and E. Witten \rpri\ have shown that there is no
perturbative superpotential generated for these moduli, thus one has
a nice compact 326 complex dimensional moduli space of vacua, containing 
singular loci of at least complex codimension one. 

Is there another $(0,2)$ model which agrees with the quintic at the 
Landau-Ginzburg point? Indeed there is, the singular configuration 
can be written as
\eqn\qdual{ V(1,1,1,2;5) \to \IP_{1,1,1,1,1,3}[4,4].}
One new coordinate of weight $\omega=3$ was introduced, which however
generates a mass term $\lambda_4 x_6$ in the $(0,2)$ superpotential and 
therefore can be integrated out. We already know that there do not exist
any further moduli at the Landau-Ginzburg point which are not moduli
of the quintic. Therefore, there cannot be a transition from the 
quintic to this dual model. Every deformation around the quintic 
Landau-Ginzburg point corresponds to a deformation of the dual model.  
Thus, the most natural scenario is, that the two moduli spaces of the 
quintic and the dual model are isomorphic. One necessary condition for
this picture to be true is that the dual model at large radius also
has a 326 dimensional moduli space. Thus, one is facing the task of
calculating bundle valued cohomology for $(0,2)$ models. This is 
a quite technical process but I am under the impression 
that these techniques \rend, in particular
those for calculating $h^1(M,{\rm End}(V_M))$, are not so well known and 
therefore, they will be reviewed  and generalized  here.  

\subsec{Toric resolution}

To begin with, since the base manifold of the model 
\eqn\qdual{ V(1,1,1,2;5) \to \IP_{1,1,1,1,1,3}[4,4]}
contains a $\ZZ_3$ singularity one has to resolve the ambient space. 
To this end,  methods known from toric geometry are used \rvitja
\footnote{$^1$}{For some of the calculations involving toric varieties
the maple packages {\it Schubert} and {\it Puntos} have been used \rschu}. 
The vertices of the fan describing the resolved toric variety are
\eqn\vertices{\eqalign{ &v_1=(1,0,0,0,0),\ v_2=(0,1,0,0,0),\ v_3=(0,0,1,0,0),\ 
               v_4=(0,0,0,1,0), \cr
               &v_5=(0,0,0,0,1),\ v_6=(-1,0,0,0,0),\ v_7=(-3,-1,-1,-1,-1). }}
The only star subdivision  is
\eqn\triangle{\eqalign{ {\rm Cones}=\{ &[1,2,3,4,5], [1,2,3,4,7], [1,2,3,5,7],
      [1,2,4,5,7], [1,3,4,5,7], \cr  &[2,3,4,5,6], [2,3,4,6,7], 
         [2,3,5,6,7], [2,4,5,6,7], [3,4,5,6,7] \},  }} 
which yields the Stanley-Reisner ideal 
\eqn\srideal{ {\rm SR}=\{ x_1\, x_6,\  x_2\, x_3\, x_4\, x_5\, x_7\} .}
The charges of the fields in the linear $\sigma$-model are given by the 
kernel of the matrix of all vertices. In our case there are two
$U(1)$s under which the fields carry charges. 
For the fields defining the base threefold these charges are presented
in Table 3.1.  
\vskip 0.1in
\meno
\cl{\vbox{
\hbox{\vbox{\offinterlineskip
\def\tablespace{height2pt&\omit&&\omit&&\omit&&\omit&&\omit&&
                          \omit&&\omit&&\omit&&\omit&\cr}
\def\tablerule{\tablespace\noalign{\hrule}\tablespace}

\hrule\halign{&\vrule#&\strut\hskip0.2cm\hfil#\hfill\hskip0.2cm\cr
\tablespace
& $x_1$ && $x_2$ && $x_3$ && $x_4$ && $x_5$ && $x_6$ && $x_7$ && 
$\Gamma_1$ && $\Gamma_2$ &\cr
\tablerule
& $1$ && $0$  && $0$ && $0$ && $0$ && $1$ && $0$  &&  $-1$ && $-1$ &\cr
& $3$ && $1$  && $1$ && $1$ && $1$ && $0$ && $1$  &&  $-4$ && $-4$ &\cr
\tablespace}\hrule}}}}
\cl{
\hbox{{\bf Table 3.1:}{\it ~~Charges for the base.}}}
\meno
The resolution of the sheaf has to be done in such a way that all
quadratic anomaly cancellation conditions for the chiral fermions 
are satisfied and that the sheaf  agrees with the unresolved one on
the singular locus. 
In our case the resolution is given by assigning the $U(1)$ charges in 
Table 3.2 to the left moving fermions in the linear $\sigma$-model.
\vskip 0.1in
\meno
\cl{\vbox{
\hbox{\vbox{\offinterlineskip
\def\tablespace{height2pt&\omit&&\omit&&\omit&&\omit&&\omit&&
                          \omit&\cr}
\def\tablerule{\tablespace\noalign{\hrule}\tablespace}

\hrule\halign{&\vrule#&\strut\hskip0.2cm\hfil#\hfill\hskip0.2cm\cr
\tablespace
& $\lambda_1$ && $\lambda_2$ && $\lambda_3$ && $\lambda_4$ && $\lambda_5$ && 
   p  &\cr
\tablerule
& $1$ && $0$  && $0$ && $0$  && $0$  &&  $-1$  &\cr
& $0$ && $1$  && $1$ && $1$ && $2$ &&  $-5$  &\cr
\tablespace}\hrule}}}}
\cl{
\hbox{{\bf Table 3.2:}{\it ~~Charges for the bundle.}}}
\meno
In order to get a sheaf of rank three on the resolved space one 
has to introduce
one fermionic gauge symmetry so that the sheaf  is given by the
cohomology of the monad
\eqn\monad{ 0\to\ {\cal O}\vert_{M} \to {\cal O}(1,0)\oplus 
                  {\cal O}(0,1)^3\oplus{\cal O}(0,2)\vert_{M}
                  \to{\cal O}(1,5)\vert_{M}\to0.}
Denoting the two independent divisors in the ambient  space $A$ 
spanning $H_4(A)=H^2(A)$ as $\eta_1$ and 
$\eta_2$, a field with charges $(Q_1,Q_2)=(m,n)$ can be regarded as a 
section of the line bundle ${\cal O}(m\eta_1+n\eta_2)$. These line bundles
are written as ${\cal O}(m,n)$. 
In order for the sheaf  $V_M$ to be non-singular one has to choose  the maps
$E_a(x)$ and $F_a(x)$ in such a way that each set does not vanish 
simultaneously on $M$. The general form  of the hypersurfaces and 
functions $F_a$ is
\eqn\FFs{ \eqalign{ &G^{(1,2)}_{(1,4)}=x_1\, P_1(y) + x_6\, P_4(y) \cr
                    &F^1_{(0,5)}=P_5(y),\ \ F^{(2,3,4)}_{(1,4)}=
                    x_1\, P_1(y) + x_6\, P_4(y),\ \ 
                    F^5_{(1,3)}=x_1+x_6\,P_3(y). }}
For generic choice of the polynomials $P_n(y)$ with $y=(x_2,x_3,x_4,x_5,x_7)$
vanishing of the $G^j$s and $F_a$s implies either $x_1=x_6=0$ or 
$x_2=x_3=x_4=x_5=x_7=0$. However,  due to the Stanley-Reisner ideal 
both sets are excluded from the ambient toric variety.
For the $E_a$s one can take 
\eqn\EEs{ \eqalign{ E^1_{(1,0)}=x_6,\ \ E^{(2,3,4)}_{(0,1)}=P_1(y),\  \ 
                    E^5_{(0,2)}=P_2(y).}}
For this choice of data the coherent sheaf defined by the monad \monad\  is 
actually a vector bundle and we will call it this way in the following
discussion. 
Using the Stanley-Reisner ideal one first calculates the intersection ring
of the ambient toric variety
\eqn\intringa{ 81\eta_1^5 -27\eta_1^4\eta_2 + 9\eta_1^3\eta_2^2 - 
                3\eta_1^2\eta_2^3 + \eta_1\eta_2^4}
and afterwards the intersection ring of the complete intersection threefold
\eqn\intringb{ 9\eta_1^3 -3\eta_1^2\eta_2 + \eta_1\eta_2^2 + 
                5\eta_2^3.}
Splitting the monad into two exact sequences
\eqn\exseq{\eqalign{&(E):\ 0\to {\cal O}\vert_{M} \to {\cal O}(1,0)\oplus 
                  {\cal O}(0,1)^3\oplus{\cal O}(0,2)\vert_{M}
                  \to {\cal E}_M\to 0 \cr
                   &(V):\ 0\to \ V_M\to {\cal E}_M 
                  \to{\cal O}(1,5)\vert_{M}\to0,}}
it is straightforward to compute  the third Chern class of the bundle 
$V_M$. One finds $c_3(V_M)=-200$ which nicely agrees with the Landau-Ginzburg
calculation and the result for the quintic. The same computation  can be
carried out for the tangent bundle leading to the Euler number of the base 
manifold $\chi(M)=-168$. 
This is very encouraging and more refined topological
numbers will be calculated in the following two subsections. 
But before that,  the phase structure of the K\"ahler moduli space
of the resolved model is briefly discussed in order to ensure 
that there still is a Landau-Ginzburg phase. 
The $D$-terms in the linear $\si$-model are
\eqn\dterms{\eqalign{ &D_1=\vert x_1\vert^2 + \vert x_6\vert^2 - 
                          \vert p\vert^2 -r_1 \cr
                      &D_2=3\vert x_1\vert^2 + \vert x_2\vert^2 +
                       \vert x_3\vert^2 + \vert x_4\vert^2 +
                       \vert x_5\vert^2 + \vert x_7\vert^2 -
                        5\vert p\vert^2 -r_2 \cr}}
leading to the following four  phases in the K\"ahler moduli space.
\smno
{\bf Phase I:} $r_1>0$ and $r_2-3r_1>0 $ \pano
      This is the Calabi-Yau phase. The $x_i$ take values
      in the complete intersection in the toric variety and the left moving 
     fermions live on  the bundle described by the monad \monad.\smno
{\bf Phase II:} $r_2>0$ and $r_2-3r_1<0$ \pano
      In this second geometric region one has an orbifold phase. 
       The base manifold is the singular $\IP_{1,1,1,1,1,3}[4,4]$ 
       with a V bundle over it.  \smno
{\bf Phase III:} $5r_1-r_2>0$ and $r_2<0$ \pano
       This is a Landau-Ginzburg phase, where the 
               expectation values
               of $\vert p\vert^2$ and $\vert x_6\vert^2$ are fixed at 
               $-r_2/5$ and $(5r_1-r_2)/5$, respectively. The remaining 
               coordinate fields have vanishing expectation value and 
               their fluctuations are governed by a superpotential which after
            integrating out $x_1$ looks the same as that for the quintic.\smno
{\bf Phase IV:} $5r_1-r_2<0$ and $r_1<0$ \pano
     This is some hybrid phase, in which $x_1=x_6=0$, 
                    $\vert p\vert^2=-r_1$ and the remaining five $x_i$
                     take values in a quintic $F_1(y)=0$
                      with K\"ahler class $r_2-5r_1$.
                    However, since $x_1=x_6=0$ the left and right
                 moving fermions
                    do not take values in the tangent bundle of this quintic,
                   so that one does not simply get a non-linear $\si$-model.
                In this phase the $(0,2)$ model seem to memorize its
                  connection to the quintic, as well.

\subsec{Complex and K\"ahler deformations}

In general a short exact sequence of sheaves 
\eqn\ges{ 0 \to A \buildrel \alpha \over \to B \buildrel \beta \over 
        \to C \to 0}
implies a long exact sequence in cohomology
\eqn\gesc{ 0 \to H^0(M,A) \buildrel \alpha \over \to
                 H^0(M,B) \buildrel \beta \over \to
                 H^0(M,C) \buildrel \phi \over \to
                 H^1(M,A) \buildrel \alpha \over \to
                 H^1(M,B) \to\ldots.}
The maps $\alpha$ and $\beta$ in \gesc\ are induced from the sheaf
homomorphisms in \ges. For the definition of $\phi$ it is refered to
the mathematical literature \rend, but it is emphasized 
that the definition of $\phi$ relies on the shortness of
the sequence \ges. 
In order to use the long exact cohomological sequences implied by
\exseq, one  has to know the cohomology classes of line bundles restricted
to the complete intersection locus. To this end one uses the Koszul
sequence for a complete intersection of $K$ hypersurfaces
$\xi=(f_1,\ldots,f_K)$ with $f_i$ a section of the line bundle
${\cal E}_{f_i}$ over the ambient space
\eqn\kosgen{ 0\to \wedge^K {\cal E}^* \buildrel \xi\cdot\over \to \ldots
             \buildrel \xi\cdot\over \to \wedge^2 {\cal E}^*
             \buildrel \xi\cdot\over \to {\cal E}^*
             \buildrel \xi\cdot\over \to {\cal O}
             \buildrel \rho\cdot\over \to {\cal O}\vert_M \to 0 .}  
Here ${\cal E}=\bigoplus {\cal E}_{f_i}$ and ${\cal O}$ denotes 
the structure sheaf of the ambient space. 
In our case, after multiplication of \kosgen\ with a vector bundle
$T$, one  simply obtains the exact sequence
\eqn\koszul{ 0\to T\otimes {\cal O}(-2,-8) \to T\otimes \left(
             {\cal O}(-1,-4)\oplus {\cal O}(-1,-4)\right) \to
             T\to T\vert_{M}\to 0. }
Throughout the following computation $T$ will always be
a line bundle, which means that all one has to
know as input are the cohomology classes of line bundles over
the ambient toric variety. 
Luckily, by using the algorithm of \rresolv\ one can derive  closed formulas
for the dimension of these classes. Setting the binomial coefficient of
a negative number over a positive number to zero, the only non-zero 
classes are
\eqn\cohom{\eqalign{ &h^0({\cal O}(m,n))=\sum_{l=0}^m {n-3l+4\choose 4}, 
                \quad {\rm for}\ m,n\ge 0 \cr 
               &{\rm\  \breve Cech\  representative}:\quad  P(x_1,x_6,y_i) \cr
                   & \cr
                   &h^1({\cal O}(-2-m,n-3(m+1)))=
                     \sum_{l=0}^m {n-3l+4\choose 4}, 
                    \quad {\rm for}\ m,n\ge 0 \cr
                  &{\rm\  \breve Cech\ representative\ on}\  
                   \{x_1\neq 0\}\cap \{x_6\neq 0\}:\quad 
                    {P(y_i) \over x_1 x_6 Q(x_1,x_6)}  \cr
                    & \cr
                    &h^4({\cal O}(m,3m-n-5))=
                     \sum_{l=0}^m {n-3l+4\choose 4}, 
                    \quad {\rm for}\ m,n\ge 0 \cr
                  &{\rm\  \breve Cech\  representative\ on}\ 
                   {\textstyle\bigcap_i}\{y_i\neq 0\}:\quad 
                    {P(x_1,x_6) \over y_2 y_3 y_4 y_5 y_7 Q(y_i)}  \cr
                    & \cr
                    &h^5({\cal O}(-2-m,-8-n))=
                     \sum_{l=0}^m {n-3l+4\choose 4}, 
                    \quad {\rm for}\ m,n\ge 0 \cr
                  &{\rm \ \breve Cech\  representative\ on}\ 
                      {\textstyle\bigcap_i}\{y_i\neq 0\}\cap
                  \{x_1\neq 0\}\cap \{x_6\neq 0\}:\quad \cr
        &\quad\quad {1\over x_1 x_6 y_2 y_3 y_4 y_5 y_7 Q(x_1,x_6,y_i)}. \cr}}
                     
It is checked in many examples that these numbers are consistent
with the Euler characteristic $\chi(A,{\cal O}(m,n))$ of a line bundle
over the ambient space as determined  by the Riemann-Roch-Hirzebruch 
theorem
\eqn\rrht{ \chi(A,{\cal O}(m,n))=\sum_{q=0}^{{\rm dim}A} 
             (-)^q h^q(A,{\cal O}(m,n))=
            \int_A \left[ e^\lambda \prod_{i=1}^{N_x}
            {l_i \over 1-e^{-l_i} } \right]. }
In \rrht\ $\lambda$  is the first Chern class of the line 
bundle ${\cal O}(m,n)$ and the $l_i$ denote the first Chern classes 
of the homogeneous coordinates defining the toric variety.
Plugging in all the data from Table 3.1, one obtains for the Euler 
characteristic
\eqn\eul{\eqalign{ \chi(A,{\cal O}(m,n))&={1\over 120}
    (1 + m)(120 - 126m + 201m^2 - 216m^3 + 81m^4 + 250n - 
              300mn+ \cr
           &315m^2n - 135m^3n + 175n^2 - 180mn^2 + 
              90m^2n^2 + 50n^3 -  30mn^3 + 5n^4).\cr}}

As already mentioned, the two short exact sequences \exseq\ imply two 
long exact sequences of the 
cohomology groups. If these long exact sequences contain enough zeros
then one can hope to deduce the cohomology classes of the bundle
$V_M$ without a detailed study of the maps in these sequences. The Koszul 
sequence \koszul\ does not simply imply one exact sequence in cohomology 
but instead gives rise to  a spectral sequence or alternatively to 
the following three short exact 
sequences with their associated long cohomological sequences. 
\eqn\kosthree{ \eqalign{ &0\to T\otimes {\cal O}(-1,-4) \to T \to T\vert_{N}
                           \to 0 \cr
                          &0\to T\otimes {\cal O}(-2,-8) \to 
                        T\otimes {\cal O}(-1,-4) \to 
                        T\otimes {\cal O}(-1,-4)\vert_{N}
                           \to 0 \cr
                        &0\to T\otimes {\cal O}(-1,-4)\vert_{N} \to
                         T\vert_{N} \to T\vert_{M}\to 0, \cr }}
where $N$ denotes one of the two hypersurfaces, $M\subset N\subset A$.

For determining the generations and antigenerations, using 
\kosthree\   one first has to
calculate the cohomology groups on $M$ listed in Table 3.3. 
\vskip 0.1in
\meno
\cl{\vbox{
\hbox{\vbox{\offinterlineskip
\def\tablespace{height2pt&\omit&&\omit&&\omit&&\omit&&\omit&&
                          \omit&\cr}
\def\tablerule{\tablespace\noalign{\hrule}\tablespace}

\hrule\halign{&\vrule#&\strut\hskip0.2cm\hfil#\hfill\hskip0.2cm\cr
\tablespace
& $$ && ${\cal O}\vert_{M}$ && ${\cal O}(1,0)\vert_{M}$ && 
   ${\cal O}(0,1)\vert_{M}$ && ${\cal O}(0,2)\vert_{M}$ && 
   ${\cal O}(1,5)\vert_{M}$ &\cr
\tablerule
& $h^0$ && $1$  && $1$ && $5$  && $15$  &&  $131$  &\cr
& $h^1$ && $0$  && $0$ && $0$ && $0$ &&  $0$  &\cr
& $h^2$ && $0$  && $0$ && $0$  && $0$  &&  $0$  &\cr
& $h^3$ && $1$  && $0$ && $0$ && $0$ &&  $0$  &\cr
\tablespace}\hrule}}}}
\cl{
\hbox{{\bf Table 3.3:}{\it ~~Cohomology of  line bundles on $M$ .}}}
\meno
Then the first sequence in \exseq\ implies the cohomology of ${\cal E}_M$ to
be $h({\cal E}_M)=(h^0,h^1,h^2,h^3)=(30,0,1,0)$. From the second sequence 
one obtains $h(V_M)=(0,101,1,0)$, thus the model
has $101$ generations in the ${\bf 27}$ representation of $E_6$ and
$1$ antigeneration in the $\overline{\bf 27}$ representation of $E_6$.

Carrying out the analogous computation for the tangent bundle of
the base space, which is given by the two short exact sequences
\eqn\exseqt{\eqalign{&0\to {\cal O}\oplus {\cal O}  \vert_{M} \to 
                  {\cal O}(1,0)\oplus 
                  {\cal O}(0,1)^5\oplus{\cal O}(1,3)\vert_{M}
                  \to {\cal F}_M\to 0 \cr
                   &0\to T_M\to {\cal F}_M 
                  \to{\cal O}(1,4)\oplus {\cal O}(1,4) \vert_{M}\to0,}}
one gets $h(T_M)=(0,86,2,0)$. Thus the base manifold has $86$ complex 
deformations and $2$ K\"ahler deformations. 
However, this is only part of the large radius moduli space of heterotic
string compactifications. The remaining part are the bundle deformations
parameterized by elements in $H^1(M,\rm{End}(V_M))$.

\subsec{Bundle deformations}

The bundle endomorphisms $\rm{End}(V_M)$ are by definition the traceless
part of the bundle $V_M\otimes V_M^*$. Using $tr(V_M\otimes V_M^*)=
{\cal O}\vert_M$ and that for stable bundles on a Calabi-Yau $n$-fold
the lowest and highest cohomology groups $H^0$ and $H^n$ vanish,
one obtains the following relation between $h^q(M,\rm{End}(V_M))$ 
and $h^q(M,V\otimes V^*)$ \rend
\eqn\stable{ h^q(M,V_M\otimes V_M^*)=\cases{ 1 \quad &{\rm for}\ q=0,3\cr
                             h^q(M,\rm{End}(V_M)) \quad &{\rm for}\ q=1,2.\cr}}
Furthermore, Serre duality implies 
$h^1(M,V_M\otimes V_M^*)=h^2(M,V\otimes V^*)$
which will provide a non-trivial check whether our calculation is correct.

To begin with, the exact sequence \exseq(V) is dualized and tensored 
with $V_M$ yielding the exact sequence
\eqn\enda{ 0 \to V_M\otimes {\cal O}(-1,-5)\vert_{M} \to
             V_M\otimes {\cal E}^*_M \to V_M\otimes V_M^* \to 0 }
containing the desired bundle. 
\pano
$\bullet$  $V_M\otimes {\cal O}(-1,-5)\vert_{M}$
\pano
To determine this bundle the following two exact sequences are used
which are derived from \exseq. 
\eqn\endb{\eqalign{ 0 \to &V_M\otimes {\cal O}(-1,-5)\vert_{M} \to
                     {\cal E}_M\otimes {\cal O}(-1,-5)\vert_{M} \to
                     {\cal O}\vert_{M}\to 0 \cr
                     0\to &{\cal O}(-1,-5)\vert_{M}\to
                      {\cal O}(0,-5)\oplus {\cal O}(-1,-4)^3\oplus
                      {\cal O}(-1,-3)\vert_{M}\to \cr
                     & {\cal E}_M\otimes {\cal O}(-1,-5)\vert_{M} \to 0.}}
Again one uses the Koszul sequence to determine the cohomology classes
of all involved line bundles restricted to $M$ and finally gets the results
displayed in Table 3.4.
\vskip 0.1in
\meno
\cl{\vbox{
\hbox{\vbox{\offinterlineskip
\def\tablespace{height2pt&\omit&&\omit&&\omit&&
                          \omit&\cr}
\def\tablerule{\tablespace\noalign{\hrule}\tablespace}

\hrule\halign{&\vrule#&\strut\hskip0.2cm\hfil#\hfill\hskip0.2cm\cr
\tablespace
& $$ && $V_M\otimes {\cal O}(-1,-5)\vert_{M}$ && 
        ${\cal E}_M\otimes {\cal O}(-1,-5)\vert_{M}  $ && 
   ${\cal O}\vert_{M} $ & \cr
\tablerule
& $h^0$ && $0$  && $0$ && $1$  &\cr
& $h^1$ && $1$  && $0$ && $0$  &\cr
& $h^2$ && $0$  && $0$ && $0$  &\cr
& $h^3$ && $248$  && $249$ && $1$  &\cr
\tablespace}\hrule}}}}
\cl{
\hbox{{\bf Table 3.4:}{\it ~~determine 
 $V_M\otimes {\cal O}(-1,-5)\vert_{M}$  }}}
\meno
\pano
$\bullet$  $V_M\otimes {\cal E}_M^*$
\pano
The following sequence is used as the starting point to determine 
this bundle
\eqn\endc{ 0 \to V_M\otimes {\cal E}^*_M  \to {\cal E}_M\otimes {\cal E}^*_M
            \to {\cal E}^*_M\otimes {\cal O}(1,5)\vert_{M} \to 0.}
The cohomology of ${\cal E}^*_M\otimes {\cal O}(1,5)\vert_{M}$ is 
given by Serre duality as $h({\cal E}^*_M\otimes {\cal O}(1,5)\vert_{M})=
(249,0,0,0)$. 
\meno
\pano
$\bullet$  ${\cal E}_M\otimes {\cal E}^*_M$
\pano
Dualizing \exseq(E) and tensoring with ${\cal E}_M$ gives the exact sequence
\eqn\endd{ 0 \to {\cal E}_M\otimes {\cal E}^*_M \to {\cal E}_M\otimes\left(
            {\cal O}(-1,0)\oplus{\cal O}(0,-1)^3\oplus{\cal O}(0,-2)\right)
            \vert_{M} \to {\cal E}_M \to 0,}
which allows one to determine ${\cal E}_M\otimes {\cal E}^*_M$ after 
having found the cohomology of all the 
${\cal E}_M\otimes{\cal O}(m,n)\vert_{M}$. However, the latter can be 
computed  analogously to ${\cal E}_M\otimes {\cal O}(-1,-5)\vert_{M}$ in 
the second sequence in \endb. 
Tracing through all these sequences finally gives Table 3.5.
\vskip 0.1in
\meno
\cl{\vbox{
\hbox{\vbox{\offinterlineskip
\def\tablespace{height2pt&\omit&&\omit&&\omit&&
                          \omit&\cr}
\def\tablerule{\tablespace\noalign{\hrule}\tablespace}

\hrule\halign{&\vrule#&\strut\hskip0.2cm\hfil#\hfill\hskip0.2cm\cr
\tablespace
& $$ && ${\cal E}_M\otimes {\cal E}^*_M$ && 
        ${\cal E}_M\otimes\left(
            {\cal O}(-1,0)\oplus{\cal O}(0,-1)^3\oplus{\cal O}(0,-2)\right)
            \vert_{M}$ && 
   ${\cal E}_M$ & \cr
\tablerule
& $h^0$ && $11$  && $41 =\ 16\ +\ 24\ +\ 1$ && $30$  &\cr
& $h^1$ && $0$  && $0$ && $0$  &\cr
& $h^2$ && $0$  && $0$ && $1$  &\cr
& $h^3$ && $11$  && $10\ =\ 0\ +\ 0\ +\ 10$ && $0$  &\cr
\tablespace}\hrule}}}}
\cl{
\hbox{{\bf Table 3.5:}{\it ~~determine ${\cal E}_M\otimes {\cal E}^*_M$.}}}
\meno
Here Serre duality was used to fix the possible additive 
offset to $h^0$ and $h^1$ to be zero. 
Going back to \endc\ one obtains that $V_M\otimes {\cal E}_M^*$ is fixed 
as displayed in Table 3.6.
\vskip 0.1in
\meno
\cl{\vbox{
\hbox{\vbox{\offinterlineskip
\def\tablespace{height2pt&\omit&&\omit&&\omit&&
                          \omit&\cr}
\def\tablerule{\tablespace\noalign{\hrule}\tablespace}

\hrule\halign{&\vrule#&\strut\hskip0.2cm\hfil#\hfill\hskip0.2cm\cr
\tablespace
& $$ && $V_M\otimes {\cal E}_M^*$ && 
        ${\cal E}_M\otimes {\cal E}^*_M$ && 
   ${\cal E}_M^*\otimes{\cal O}(1,5)\vert_{M} $ & \cr
\tablerule
& $h^0$ && $x$  && $11$ && $249$  &\cr
& $h^1$ && $x+238$  && $0$ && $0$  &\cr
& $h^2$ && $0$  && $0$ && $0$  &\cr
& $h^3$ && $11$  && $11$ && $0$  &\cr
\tablespace}\hrule}}}}
\cl{
\hbox{{\bf Table 3.6:}{\it ~~determine $V_M\otimes {\cal E}_M^*$.}}}
\meno
The final step is to use \enda\ to determine $V_M\otimes V_M^*$
as shown in Table 3.7.
\vskip 0.1in
\meno
\cl{\vbox{
\hbox{\vbox{\offinterlineskip
\def\tablespace{height2pt&\omit&&\omit&&\omit&&
                          \omit&\cr}
\def\tablerule{\tablespace\noalign{\hrule}\tablespace}

\hrule\halign{&\vrule#&\strut\hskip0.2cm\hfil#\hfill\hskip0.2cm\cr
\tablespace
& $$ && $V_M\otimes {\cal O}(-1,-5)\vert_{M}$ && 
        $V_M\otimes {\cal E}_M^*$ && 
   $V_M\otimes V_M^*$ & \cr
\tablerule
& $h^0$ && $0$ && $x$  && $1$  &\cr
& $h^1$ && $1$ && $x+238$  && $238$  &\cr
& $h^2$ && $0$ && $0$  && $238$  &\cr
& $h^3$ && $248$ && $11$   && $1$  &\cr
\tablespace}\hrule}}}}
\cl{
\hbox{{\bf Table 3.7:}{\it ~~determine $V_M\otimes V_M^*$.}}}
\meno
It is assumed here that the bundle is stable and due to \stable\ 
that $h^0=h^3=1$.
Thus, it is derived that the bundle has $h^1(M,{\rm End}(V_M))=238$ 
deformations.
Adding up all the moduli at large radius one obtains  
$h^1(M,T)+h^2(M,T)+h^1(M,{\rm End}(V_M))=86+2+238=326$ which is the same number
of moduli as for the quintic. Due to this nice non-trivial matching of 
the dimensions of the moduli spaces both at large and at small radius and
the equality of the two models at their Landau-Ginzburg locus
it is conjectured that the two moduli spaces are actually the same. A physicist
doing experiments in four dimensions cannot decide whether the hidden 
six dimensional world is the quintic or the dual $(0,2)$ model.

It is clear that the mapping between the two sets of moduli must include
exchanges between the three classes of moduli. For instance, at large 
radius of the dual $(0,2)$ model the vector bundle $V_M$ is definitely 
different from the tangent bundle implying that the model is really 
$(0,2)$ and not
$(2,2)$ supersymmetric. However, the quintic has a $(2,2)$ subset even for
large radius. Thus, it cannot be that K\"ahler moduli are only mapped to 
K\"ahler moduli, instead K\"ahler moduli seem to be mapped to bundle moduli.
This picture is also consistent with the fact that at the Landau-Ginzburg 
point there exist 25 twisted singlets \rkw. For the quintic, only one of these
singlets corresponds to the K\"ahler deformation, the other 24 are bundle
deformations. For the dual $(0,2)$ model one expects two combinations of these 
24 twisted singlets to constitute the new K\"ahler deformations. It is
tempting to speculate about a maximal $(0,2)$ model dual to the quintic,
for which all 25 twisted singlets do correspond to some K\"ahler moduli.

Clearly, it would
be very interesting to gain  a better understanding of how the moduli spaces
are actually mapped to each other. Similarly to mirror
symmetry this would allow one to derive statements about some properties of
the moduli space of one model by knowing it for the dual model. For instance,
if the duality is correct,  one can already learn that the moduli space of the
$(0,2)$ model contains a $102$ dimensional sublocus of  $(2,2)$ world-sheet 
supersymmetry, which is absolutely not obvious by knowing the model 
only in the large 
radius phase. Using the results from mirror symmetry \rcogp, one even knows
the metric on the one dimensional subset of this $(2,2)$ locus 
which is mapped to the K\"ahler moduli space of the quintic. 

An interesting observation one can make is that the base manifold
of the $(0,2)$ model is related to the complete intersection manifold
\eqn\cim{ \matrix{ \IP_4 \cr \IP_1 \cr}\left[\matrix{4 & 1 \cr 1 & 1 \cr}
          \right] }
by what is called  a flip transition in \rcgh. This means that there exists a 
conifold transition from  the quintic to the base manifold of the dual 
$(0,2)$ model. The duality proposed here has at first sight nothing to
do with a singular transition and in the moment I do not know whether
the appearance of the conifold transition here is only a coincidence or
has really to teach us something about the non perturbative resolution 
of the conifold singularity in the heterotic string context.

\newsec{The Sextic and its duals}

In the last section one particular simple example was studied in very
much detail revealing that in that case there cannot be a transition 
between different moduli spaces, instead we  were  led to the conjecture 
that the two moduli spaces are isomorphic. However, the quintic is
special in the sense that at the Landau-Ginzburg locus the number of
gauge singlets matches exactly the number of moduli at large radius.
This fact excluded the possibility of a transition just from  the very 
beginning.
In this section another example will be studied, for which at the 
Landau-Ginzburg
locus the number of singlets increases. Thus, this example is closer to
the generic case than the quintic but fortunately it is still easy enough to
be treated using the techniques from the last section.  

\subsec{The Sextic}

The base manifold is a hypersurface of degree six in the weighted projective
space $\IP_{1,1,1,1,2}$ and the bundle is a deformation of the tangent bundle.
At the Landau-Ginzburg point the model has $N_{27}=103$ generations and
$N_{\overline{27}}=1$ antigenerations. Moreover, the spectrum contains
$N_1=${\bf 340} gauge singlets, with $307$, $27$ and $6$ arising from
the $k=1$, $k=3$ and $k=5$ twisted sector, respectively.  

The number of moduli in the geometric phase is  again the crucial unknown.
Since one needs the cohomology classes of line bundles in the ambient 
variety as input
for running the exact sequences, one has to worry about the $\ZZ_2$ 
singularity in the weighted projective space even though the hypersurface
avoids this singularity. 

Blowing up the singularity generates  the charges of the fields 
defining the base manifold. These are shown in Table 4.1.
\vskip 0.1in
\meno
\cl{\vbox{
\hbox{\vbox{\offinterlineskip
\def\tablespace{height2pt&\omit&&\omit&&\omit&&\omit&&\omit&&
                          \omit&&\omit&\cr}
\def\tablerule{\tablespace\noalign{\hrule}\tablespace}

\hrule\halign{&\vrule#&\strut\hskip0.2cm\hfil#\hfill\hskip0.2cm\cr
\tablespace
& $x_1$ && $x_2$ && $x_3$ && $x_4$ && $x_5$ && $x_6$ && 
$\Gamma_1$  &\cr
\tablerule
& $1$ && $0$  && $0$ && $0$ && $1$ && $0$  &&  $-2$  &\cr
& $2$ && $1$ && $1$ && $1$ && $0$ && $1$  &&  $-6$ &\cr
\tablespace}\hrule}}}}
\cl{
\hbox{{\bf Table 4.1:}{\it ~~Charges for the base.}}}
\meno
However, for the tangent bundle one obtains $\chi(M)=-200$ which is not
what one wants, but one can choose the  bundle resolution in Table 4.2 
which also agrees on the singular space $r_1\to 0$ with the tangent 
bundle of the sextic. 
\vskip 0.1in
\meno
\cl{\vbox{
\hbox{\vbox{\offinterlineskip
\def\tablespace{height2pt&\omit&&\omit&&\omit&&\omit&&\omit&&
                          \omit&\cr}
\def\tablerule{\tablespace\noalign{\hrule}\tablespace}

\hrule\halign{&\vrule#&\strut\hskip0.2cm\hfil#\hfill\hskip0.2cm\cr
\tablespace
& $\lambda_1$ && $\lambda_2$ && $\lambda_3$ && $\lambda_4$ && $\lambda_5$ && 
   p  &\cr
\tablerule
& $0$ && $0$  && $1$ && $1$  && $0$  &&  $-2$  &\cr
& $1$ && $1$  && $1$ && $1$ && $2$ &&  $-6$  &\cr
\tablespace}\hrule}}}}
\cl{
\hbox{{\bf Table 4.2:}{\it ~~Charges for the bundle.}}}
\meno
Indeed, this model has $c_3(V_M)=-204$. Now one has to go all the way through 
the exact sequence calculation. One finds that there are $h^1(M,V_M)=103$
generations and $h^2(M,V_M)=1$ antigenerations. The base manifold has
$h^1(M,T)=102$ complex deformations and $h^2(M,T)=2$ K\"ahler deformations. 
For the number of bundle deformations one obtains $h^1(M,{\rm End}(V_M))=234$,
so that the total number of large radius moduli is {\bf 338}. 

\subsec{Dual model A}

By introducing one new coordinate of weight $\omega=4$ and exchanging 
some of the $F_a$s with some of the hypersurface constraints $G_j$ one 
obtains the singular model
\eqn\sduala{ V(1,1,2,2;6) \to \IP_{1,1,1,1,2,4}[5,5],}
which has the same Landau-Ginzburg potential as the $(0,2)$ sextic. 
The resolution of the $\ZZ_2$ and $\ZZ_4$ singularity leads to the
charges for the base manifold displayed in Table 4.3.
\vskip 0.1in
\meno
\cl{\vbox{
\hbox{\vbox{\offinterlineskip
\def\tablespace{height2pt&\omit&&\omit&&\omit&&\omit&&\omit&&\omit&&
                          \omit&&\omit&&\omit&&\omit&\cr}
\def\tablerule{\tablespace\noalign{\hrule}\tablespace}

\hrule\halign{&\vrule#&\strut\hskip0.2cm\hfil#\hfill\hskip0.2cm\cr
\tablespace
& $x_1$ && $x_2$ && $x_3$ && $x_4$ && $x_5$ && $x_6$ && $x_7$ && $x_8$ &&
$\Gamma_1$ && $\Gamma_2$ &\cr
\tablerule
& $1$ && $0$  && $0$ && $0$ && $0$ && $1$ && $0$  && $0$ &&  $-1$ && $-1$ &\cr
& $2$ && $1$  && $0$ && $0$ && $0$ && $0$ && $1$  && $0$ &&  $-2$ && $-2$ &\cr
& $4$ && $2$  && $1$ && $1$ && $1$ && $0$ && $0$  && $1$ &&  $-5$ && $-5$ &\cr
\tablespace}\hrule}}}}
\cl{
\hbox{{\bf Table 4.3:}{\it ~~Charges for the base.}}}
\meno
The bundle resolution leads to the charges of the left moving
fermions in Table 4.4.
\vskip 0.1in
\meno
\cl{\vbox{
\hbox{\vbox{\offinterlineskip
\def\tablespace{height2pt&\omit&&\omit&&\omit&&\omit&&\omit&&
                          \omit&\cr}
\def\tablerule{\tablespace\noalign{\hrule}\tablespace}

\hrule\halign{&\vrule#&\strut\hskip0.2cm\hfil#\hfill\hskip0.2cm\cr
\tablespace
& $\lambda_1$ && $\lambda_2$ && $\lambda_3$ && $\lambda_4$ && $\lambda_5$ && 
   p  &\cr
\tablerule
& $1$ && $0$  && $0$ && $0$  && $0$  &&  $-1$  &\cr
& $0$ && $1$  && $1$ && $0$  && $0$  &&  $-2$  &\cr
& $0$ && $1$  && $1$ && $2$ && $2$ &&  $-6$  &\cr
\tablespace}\hrule}}}}
\cl{
\hbox{{\bf Table 4.4:}{\it ~~Charges for the bundle.}}}
\meno
Using the Stanley-Reisner ideal 
\eqn\srideala{ {\rm SR}=\{ x_1\, x_6,\  x_2\, x_7,\ x_3\, x_4\, x_5\, x_8\} ,}
the intersection ring on the threefold is 
\eqn\intringa{ 8\eta_1^3-2\eta_1\eta_2^2+8\eta_2^3 - 2\eta_1^2\eta_3 +
               \eta_1\eta_2\eta_3 - 4\eta_2^2\eta_3 + 2\eta_2\eta_3^2 + 
               2\eta_3^3.}
The third Chern class of the bundle and the Euler 
characteristic of the base manifold are  $c_3(V_M)=-204$ and $\chi(M)=-176$,
respectively. The exact sequence calculation reveals that the 
model has $h^1(M,V_M)=103$ generations, $h^2(M,V_M)=1$ antigenerations,
$h^1(M,T_M)=91$ complex deformations and $h^2(M,T_M)=3$ K\"ahler deformations.
After another lengthy calculation one gets $h^1(M,{\rm End}(V_M))=244$ 
bundle deformations, so that  the
total number of moduli comes out to be {\bf 338}, the same as for the sextic. 
Thus, even though in this case the Landau-Ginzburg model has two more
singlets than the geometric models, the number of moduli in the geometric 
phases agree completely! This surprising coincidence is considered as a sign 
that also in this case there is no transition between different moduli
spaces but rather an isomorphy between them. 

\subsec{Dual model B}

The duality between the sextic and model A can even be extended to a triality.
The singular model
\eqn\sdualb{ V(1,1,1,3;6) \to \IP_{1,1,1,1,2,3}[5,4]}
agrees at its Landau-Ginzburg locus with the sextic and dual model A. 
The resolution of the base and the bundle results in the assignment
of charges shown in Table 4.5 and Table 4.6, respectively.
\vskip 0.1in
\meno
\cl{\vbox{
\hbox{\vbox{\offinterlineskip
\def\tablespace{height2pt&\omit&&\omit&&\omit&&\omit&&\omit&&\omit&&
                          \omit&&\omit&&\omit&&\omit&\cr}
\def\tablerule{\tablespace\noalign{\hrule}\tablespace}

\hrule\halign{&\vrule#&\strut\hskip0.2cm\hfil#\hfill\hskip0.2cm\cr
\tablespace
& $x_1$ && $x_2$ && $x_3$ && $x_4$ && $x_5$ && $x_6$ && $x_7$ && $x_8$ &&
$\Gamma_1$ && $\Gamma_2$ &\cr
\tablerule
& $1$ && $0$  && $0$ && $0$ && $0$ && $1$ && $0$  && $0$ &&  $-1$ && $-1$ &\cr
& $2$ && $1$  && $0$ && $0$ && $0$ && $0$ && $1$  && $0$&&  $-2$ && $-2$ &\cr
& $3$ && $2$  && $1$ && $1$ && $1$ && $0$ && $0$  && $1$&&  $-5$ && $-4$ &\cr
\tablespace}\hrule}}}}
\cl{
\hbox{{\bf Table 4.5:}{\it ~~Charges for the base.}}}
\meno
\vskip 0.1in
\meno
\cl{\vbox{
\hbox{\vbox{\offinterlineskip
\def\tablespace{height2pt&\omit&&\omit&&\omit&&\omit&&\omit&&
                          \omit&\cr}
\def\tablerule{\tablespace\noalign{\hrule}\tablespace}

\hrule\halign{&\vrule#&\strut\hskip0.2cm\hfil#\hfill\hskip0.2cm\cr
\tablespace
& $\lambda_1$ && $\lambda_2$ && $\lambda_3$ && $\lambda_4$ && $\lambda_5$ && 
   p  &\cr
\tablerule
& $1$ && $0$  && $0$ && $0$  && $0$  &&  $-1$  &\cr
& $0$ && $1$  && $1$ && $0$  && $0$  &&  $-2$  &\cr
& $0$ && $1$  && $1$ && $1$ && $3$ &&  $-6$  &\cr
\tablespace}\hrule}}}}
\cl{
\hbox{{\bf Table 4.6:}{\it ~~Charges for the bundle.}}}
\meno
Here one obtains $c_3(V_M)=-204$ and $\chi(M)=-160$. The more refined 
exact sequence calculation gives $h^1(M,V_M)=103$ generations, 
$h^2(M,V_M)=1$ antigenerations,
$h^1(M,T_M)=83$ complex deformations and $h^2(M,T_M)=3$ K\"ahler deformations.
Furthermore, the model has $h^1(M,{\rm End}(V_M))=252$ bundle deformations 
so that the total number of moduli is again {\bf 338}.

\newsec{Conclusion}

It is clear that many more dual models could be studied in a similar way.
For instance,  applying an exchange of $G_j$ and $F_a$ constraints 
to the model $\IP_{1,1,1,2,2}[7]$, one obtains the
following three singular dual candidates
\eqn\quadro{\eqalign{  &V(1,1,1,4;7) \to \IP_{1,1,1,2,2,3}[5,5] \cr
                       &V(1,1,2,3;7) \to \IP_{1,1,1,2,2,4}[6,5] \cr
                       &V(1,2,2,2;7) \to \IP_{1,1,1,2,2,5}[6,6], \cr}}
which would lead to four models with isomorphic moduli spaces. 

Throughout this paper, we have restricted ourselves to coherent sheaves,
which actually are  vector bundles. As nicely shown in \rresolv,
perturbative string theory can very well live with some mild singularities
in the  bundle leading to reflexive  or torsion free sheaves.
It would be interesting whether the proposed duality extends to
this more general case. 

It might be possible to find a dual pair for which both base 
manifolds are elliptically fibered. In this case the proposed duality
should have an analogue in the F-theory dual picture. 
It might also be that this duality of $(0,2)$ compactifications  is not only 
limited to models which allow a Landau-Ginzburg description, but extended
to more general $(0,2)$ models. 
 
To summarize, by  performing an exact cohomology calculation, it was  shown
that not only at small radius but also at large radius the dimensions of the
moduli spaces of the quintic and its dual $(0,2)$ model agree. This matching
generalizes even to the case when at the Landau-Ginzburg locus
additional singlets appear. 
This result was viewed as strong indication that the
moduli spaces of the models involved are in fact isomorphic and one is
actually dealing with a perturbative target space duality.

\bigbreak\bigskip\bigskip\centerline{{\bf Acknowledgments}}\nobreak

It is a pleasure to thank Michael Flohr, Peter  Mayr and Andreas Wi\ss kirchen
for discussion.
This work is supported by NSF grant PHY--9513835.

\vfill\eject

\listrefs
\bye